# Modified approach for accounting for dissipation in theoretical description of fusion of complex nuclei

**I. I. Gontchar, M. V. Chushnyakova**
*Omsk, Russia*

e-mail: vigichar@hotmail.com

The process of fusion of complex nuclei is of significant interest as an example of the collective nuclear motion of large amplitude as well as a route for synthesis of new superheavy chemical elements. This process is accompanied by the dissipation of the energy of collective motion, at least at the last stage. The dissipative nature of fusion is accounted for in many theoretical approaches. In the present work, we propose a modified method for accounting for dissipation. The main idea is to use the superfluid model for evaluating nuclear temperature, not the Fermi-gas model like in previous approaches. The calculations of the fusion (capture) cross sections are performed for reaction 16O+92Zr at the collision energies ranging from 46 up to 70 MeV. For this reaction and at these conditions it turns out, that the more complicated superfluid model results in the cross sections which are indistinguishable from the ones obtained using the much simpler Fermi-gas model.

## Introduction

The process of fusion of complex atomic nuclei is interesting both as a route for synthesis of new superheavy chemical elements and isotopes [1] and as an example of the large amplitude collective nuclear motion [2]. This process inevitably has dissipative nature because finally part of kinetic energy of the relative motion of the nuclei goes over into the intrinsic excitation energy. Dissipative nature of the fusion process is accounted for in some methods of its theoretical modeling. Within the trajectory models (see, e.g., [3, 4, 5]) as well as in the quantum diffusion approach [2, 6] dissipation is considered explicitly. In the quantum single barrier penetration model and within the coupled channels method [7, 8, 9] dissipation is accounted for implicitly via the absorbing border: the incoming wave having crossed the barrier is not reflected. There is limited number of works in the literature where the efforts were undertaken to combine dissipation with quantum mechanics using a phenomenological Hamiltonian [10, 11].

In our previous works devoted to this topic [12, 13, 14] the trajectory model with surface friction (TMSF) have been developed which is qualitatively close to Refs. [3, 15, 16]. In the present work we focus on the method used for calculating the temperature within the TMSF. The temperature is important quantity defining the amplitude of fluctuations of the collective linear momentum corresponding to the center-to-center distance.

## 1. Trajectory model with surface friction (TMSF)

Within the TMSF, the nucleus-nucleus collision is described as motion of a fictious Brownian particle under the influence of conservative, dissipative, and random forces:

$$dp = -\frac{dU_{tot}}{dq}dt + \frac{\hbar^2 L^2}{m_q q^3}dt - \frac{p}{m_q}K_R\left[\frac{dU_n}{dq}\right]^2 dt + f\left|\frac{dU_n}{dq}\right|dW\sqrt{2\theta K_R}, \quad dq = \frac{pdt}{m_q}. \quad (1,2)$$

Here $p$ stands for the momentum corresponding to the radial motion of colliding nuclei, i.e. the linear momentum of the fictious Brownian particle modeling this motion, $[p]$=MeV·zs (1 zs=$10^{-21}$ s); $U_{tot} = U_n + U_C$ denotes the total energy of the nucleus-nucleus interaction; $U_n$ is the Strong nucleus-nucleus Potential (SnnP); $U_C$ is the Coulomb interaction energy; $K_R$ denotes the strength of the friction coefficient, $[K_R] = zs/GeV$; $\hbar L$ stands for the orbital angular momentum; $\theta$ denotes the temperature which below is described in detail, $[\theta]$=MeV. The random force (the last term in Eq. (1)) is proportional to the increment $dW$ of the Wiener process $W$; the average of this increment is equal to zero and its variance is equal to $dt$. The coefficient $f$ in Eq. (1) enables us to study the motion both in the absence of fluctuations ($f = 0$) and in the presence of those ($f = 1$).

The dimensionless generalized coordinate $q$ is related to the center-to-center distance $R$ as follows:

$$q = \frac{R}{R_P + R_T}. \qquad (3)$$

Here $R_P = r_0 A_P^{1/3}$ and $R_T = r_0 A_T^{1/3}$ are the radii of colliding nuclei which in the present work are assumed to be spherical; $r_0$=1.2 fm. The inertia parameter $m_q$ ($[m_q]$= MeV·zs$^2$) reads

$$m_q = \frac{m_n A_P A_T (R_P + R_T)^2}{A_P + A_T}. \qquad (4)$$

Here $m_n$ stands for the nucleon mass; $A_P$ ($A_T$) is the mass number of the projectile (target) nucleus.

The SnnP is calculated within the framework of the semimicroscopic double-folding model using the published computer code DFMSPH22 [17, 18] with the relativistic effective nucleon-nucleon (NN) forces [19] corresponding to the HS-parametrization [20] with the amplitude of the exchange forces 592 MeV·fm$^3$. Note, that these relativistic NN-forces result in the Coulomb barriers which are rather close to the ones obtained by means of the M3Y NN-forces (see Fig. 4 of Ref. [19]). The nucleon densities required for the double-folding calculations are taken from [14]. Eqs. (1), (2) are solved by the Runge-Kutta method of 4-th order (see details in [21, 22]).

## 2. Dissipation and fluctuations

In our former works [12-14, 21] the temperature $\theta$ was considered to be the same for both nuclei-reagents and was calculated according to the Fermi-Gas Model (FGM):

$$\theta = \sqrt{E_{DT}\left(a_1 A_T + a_2 A_T^{2/3}\right)^{-1}} = \sqrt{E_{DP}\left(a_1 A_P + a_2 A_P^{2/3}\right)^{-1}}. \qquad (5)$$

Here $a_1$=0.73 MeV$^{-1}$, $a_2$=0.095 MeV$^{-1}$ [23]. The dissipated energy $E_D$ was calculated from the energy balance:

$$E_D = E_{DP} + E_{DT} = E_{cm} - \frac{p^2}{2m_q} - U_{tot} - \frac{\hbar^2 L^2}{2m_q q^2}. \qquad (6)$$

In Eqs. (5), (6) $E_{DP}$ and $E_{DT}$ denote the intrinsic energies of the projectile- (P) and target- (T) nuclei, $E_{cm}$ is the collision energy in the center-of-mass frame.

As we see presently, a drawback of this approach is that the discrete structure of nuclear excitation is ignored. This is justified for heavy and/or deformed nuclei where the low-lying excited states possess the energies of 1 MeV or less. However, ignoring this discrete structure is hardly realistic for light and/or magic nuclei. For the nucleus $^{16}$O which is often used as a projectile in the fusion experiments, the energy of first excited state is especially high (6.05 MeV). Clearly, before the transition of a nucleus into its first excited state, the use of notion "temperature" is doubtful. The higher the energy of this state the smaller the probability of this transition. Therefore, we recon reasonable first to see what changes when we ignore the heating of the projectile. Subsequently, whole dissipated energy is concentrated in the targe nucleus ($E_D = E_{DT}$). Analysis of Eqs. (5), (6) shows that in this case the temperature rises which should result in an increase of the intensity of fluctuations in the relative motion of the colliding nuclei.

Let us now see to what extent the fluctuations are significant for the capture cross sections calculated within the TMSF. For this aim we show in Fig. 1 the ratio of the calculated cross sections $\sigma_{th}$ to the experimental ones $\sigma_{exp}$:

$$\xi_\sigma = \frac{\sigma_{th}}{\sigma_{exp}}. \qquad (7)$$

All calculations in the present work are performed for reaction $^{16}$O+$^{92}$Zr, the experimental cross sections are taken from [24, 25]. The s-wave barrier energy is $U_{B0}$= 41.9 MeV.

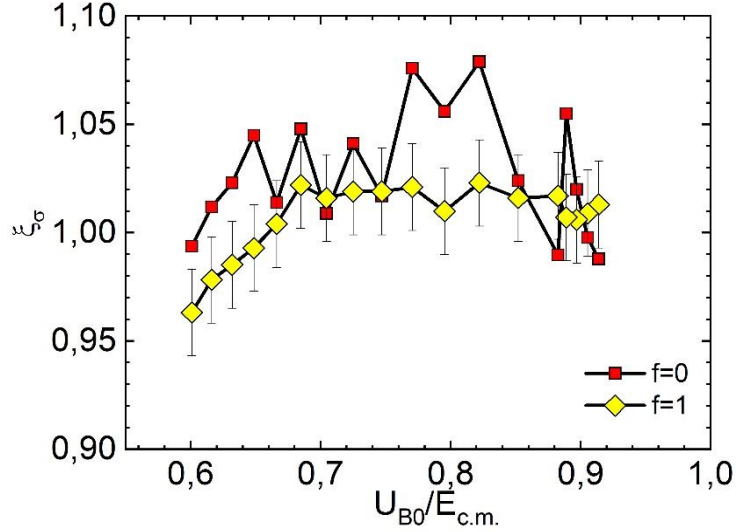

Fig.1. The ratio of the calculated cross section $\sigma_{th}$ to the experimental one $\sigma_{exp}$ (see Eq. (7)) versus the ratio of the s-wave barrier energy to the collision energy. The standard calculations accounting for fluctuations ($f = 1$) are shown by diamonds; calculations without fluctuations ($f = 0$) are shown by squares. $K_R = 17\ zs/GeV$. The dissipated energy and temperature are calculated according to Eqs. (5), (6).

The capture cross sections are calculated using the standard quantum-mechanical formula accounting for quantization of the angular momentum

$$\sigma_{th} = \frac{\pi \hbar^2}{2 m_R E_{cm}} \sum_{L=0}^{L_{max}} (2L + 1) D_L. \qquad (8)$$

Here $m_R$ denotes the reduced mass of colliding nuclei; $D_L$ is the transmission coefficient for the partial wave with the orbital quantum number $L=0,1,2,...$; $L_{max}$ stands for the maximal value of $L$ at which $D_L$ becomes negligibly small. The transmission coefficient is evaluated as the ratio of the number of trajectories for which, at given $L$, the capture conditions [12] are fulfilled to the total number of trajectories modeled for the given $L$.

In Fig. 1 the standard calculations accounting for fluctuations ($f = 1$, diamonds) are compared with the calculations without fluctuations ($f = 0$). It is seen that fluctuations sometimes reduce the calculated cross section visibly bringing it, at the given $K_R = 17\ zs/GeV$, into agreement with the experimental value. Note that earlier the role of fluctuations has been studied within the TMSF only in comparison with the memory effects [21].

As the next step we consider what happens if the projectile stays in its ground state and consequently does not absorb any dissipated energy. We mentioned already that in such calculation the temperature $\theta_T$ is expected to exceed the temperature in the "standard" approach $\theta_{PT}$ when both projectile and target nuclei share $E_D$. Results of these calculations are illustrated by Figs. 2, 3, 4. In Fig. 2a the time dependence of the temperature along one trajectory with switched off fluctuations ($f = 0$) is shown. Diamonds denote the standard calculations using Eqs. (5), (6), i.e. with $\theta_{PT}$. This temperature indeed is smaller than $\theta_T$ (solid line without symbols). The temperature $\theta_T$ is obtained when $E_{DP} = 0$, $E_{DT} = E_D$ and the second equality in Eq. (5) is not fulfilled. At the beginning of the trajectory the nuclei are far from each other and $\theta_T = \theta_{PT} = 0$ because dissipation almost absents. In Fig. 2b) we display the ratio $\theta_T / \theta_{PT}$ as the function of time for the same two trajectories as in Fig. 2a. This ratio is equal to 1.1, i.e. considering the projectile nucleus as inert one results in 10% increase of the temperature.

In Fig. 3 the same time dependences are shown as in Fig. 2, but now accounting for fluctuations. On the average the ratio $\theta_T / \theta_{PT}$ is the same in both deterministic and stochastic calculations although in the latter case the fluctuations are clearly seen.

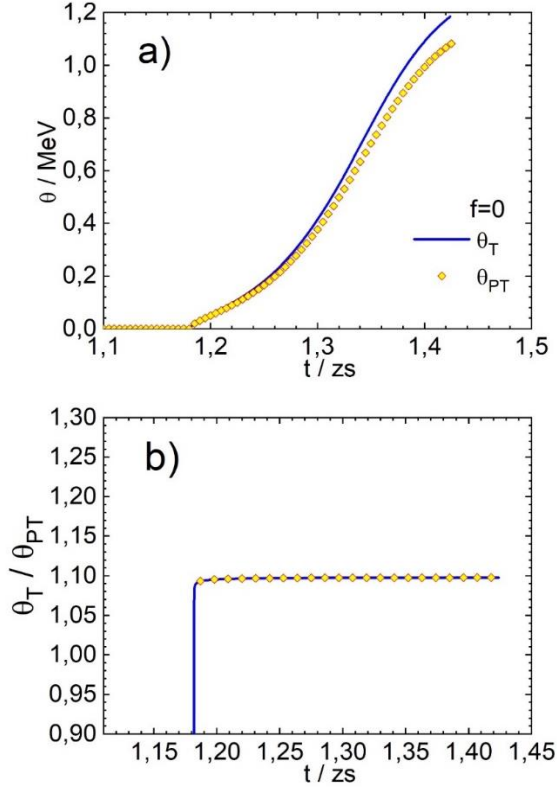 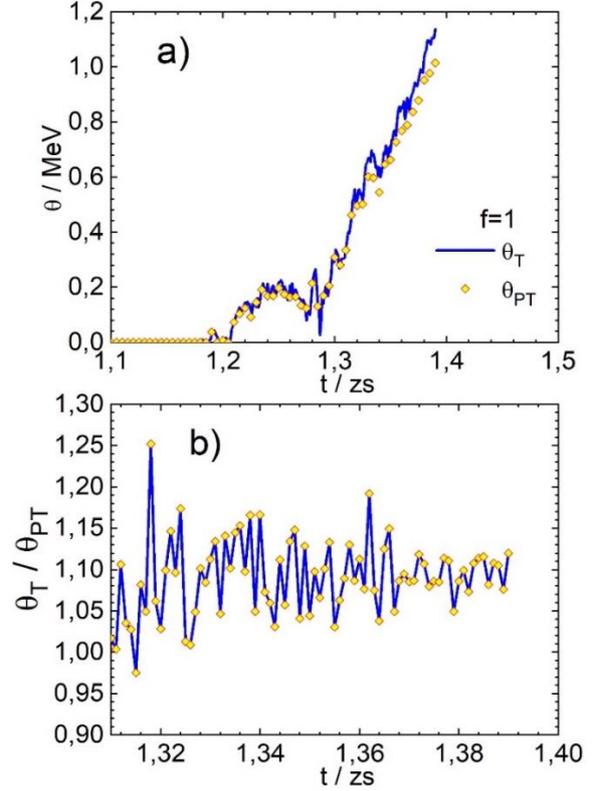

Fig. 2. a) The time dependence of the temperature along the trajectory. Diamonds ($\theta_{PT}$) correspond to the standard calculation using Eqs. (5), (6). Solid line without symbols ($\theta_T$) corresponds to the calculations with $E_{DP} = 0$, $E_{DT} = E_D$. b) The time dependence of the ratio $\theta_T/\theta_{PT}$ for the same two trajectories. Here fluctuations are switched off ($f = 0$), $E_{cm}$=51 MeV, $K_R = 17\ zs/GeV$, $L = 2$.

Fig. 3. Same as in Fig. 2 but accounting for fluctuations ($f = 1$).

The impact of the variation of the temperature ($\theta_T$ or $\theta_{PT}$) on the capture cross sections is illustrated by Fig. 4. Comparing Fig. 4 with Fig. 1 one sees that there is no impact within the statistical errors which are about 2%.

### 3. Pairs breaking and superfluid model of nucleus

The second disadvantage of the use of the Fermi-gas model for calculating the temperature within the TMSF (see Eq. (5)) is the direct use of the dissipated energy for evaluating the temperature. However, it is well known in the literature [23] that part of the excitation energy of an even-even nucleus is spent for pair breaking, and the temperature is defined by the so-called effective excitation energy

$$E_D^* = E_D - \delta_N - \delta_Z. \quad (9)$$

Here $\delta_N$ and $\delta_Z$ denote the phenomenological corrections to the nuclear binding energy. Below in the present work we assume $\delta_N + \delta_Z = 12/\sqrt{A}$ MeV according to [26].

The third disadvantage of the standard approach to the temperature calculation within the TMSF is that Eq. (5) for the temperature, i.e., the FGM, is used at extremely low values of the dissipated energy $E_D < 1$ MeV although it is known in the literature [23] that at small excitation energies (approximately smaller than 5 MeV) the nuclear superfluid model is more suitable.

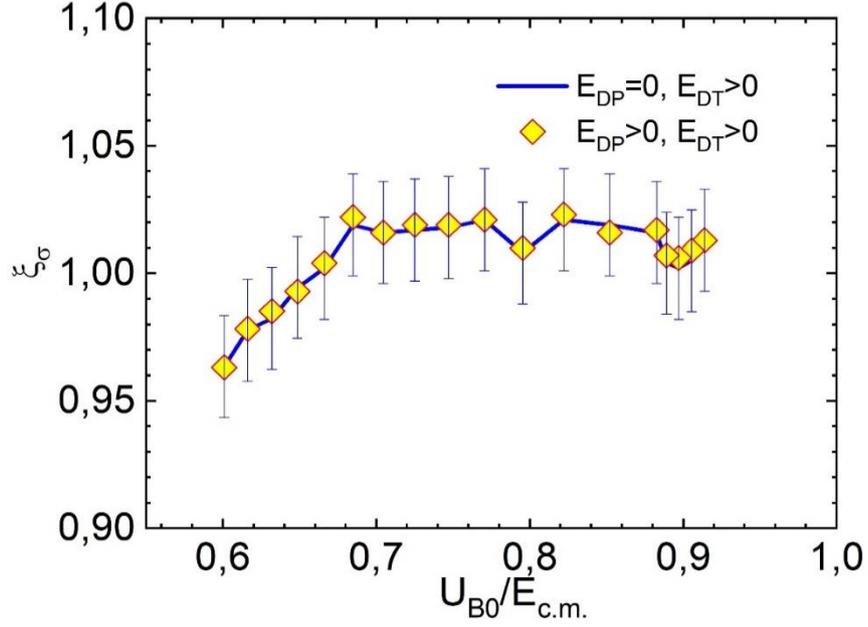

Fig. 4. The ratio of the cross sections (see Eq. (7)) versus the ratio of the s-wave barrier energy to the collision energy. Diamonds ($\theta_{PT}$) denote the standard calculation according to Eqs. (5), (6). The curve without symbols ($\theta_T$) shows the calculations with $E_{DP} = 0$, $E_{DT} = E_D$. $K_R = 17\ zs/GeV$, $f = 1$.

Within the superfluid model assuming the continuous single particle spectrum, the energy gap $\Delta$, entropy $S$, and excitation energy $E$ depend upon the temperature $\theta$. Finding $\Delta(\theta)$-dependence is the most challenging. First, one should go over to the dimensionless variables:

$$\Lambda = \frac{\Delta}{\Delta_0}, \qquad \tau = \frac{\theta}{\theta_c}. \tag{10,11}$$

Here $\Delta_0$ is the gap at zero temperature

$$\Delta_0 = \frac{12\ MeV}{\sqrt{A}}. \tag{12}$$

The critical temperature $\theta_c$ is related with $\Delta_0$ as follows

$$\theta_c = \frac{\gamma \Delta_0}{\pi} = 0.567\Delta_0. \tag{13}$$

Here $\ln \gamma = 0.577$ is the Euler constant.

Equations related the energy gap and the temperature at $0 < \theta \leq \theta_c$ can be found in [23, 27]. From the technical point of view, it is more convenient to employ different formulas for different temperature ranges:

$$\Lambda = \begin{cases} 1 - (2\gamma\tau)^{1/2} \exp\left(-\frac{\pi}{\gamma\tau}\right) & at\ 0 < \tau < 0.27, \quad (14a) \\ I_e(u) & at\ 0.27 < \tau < 0.89, \quad (14b) \\ 2\gamma \left[\frac{2(1-\tau)}{7 \cdot \zeta(3)}\right]^{1/2} & at\ 0.89 < \tau < 1; \quad (14c) \end{cases}$$

Here

$$I_e(u) = \exp\left\{-2 \int_0^\infty \frac{dx}{\sqrt{x^2 + u^2}\left[\exp(\sqrt{x^2 + u^2}) + 1\right]}\right\}. \tag{15}$$

In Eqs. (14), (15)

$$u = \frac{\Delta}{\theta} = \frac{\pi\Lambda}{\gamma\tau},\qquad(16)$$

$\zeta(3) = 1.202$ is the Riemann zeta function. Obviously, at $\tau > 1$ the gap in the elementary excitation spectrum disappears ($\Lambda=0$).

The calculation of the correlation function at very low temperature and near the transition point is relatively simple using Eqs. (14a,c). However, at the intermediate values of temperature one is forced to solve numerically the complicated Eq. (14b) which is not resolved with respect to $\Lambda$. Fig 5 illustrates the appearing problems.

After finding $\Lambda$, one evaluates the entropy as follows:

$$S = \begin{cases} \dfrac{12a\Lambda^2\theta_c}{\gamma^2\tau}\sum_{n=1}^{\infty}(-1)^{n+1}K_2\left(\dfrac{n\pi\Lambda}{\gamma\tau}\right) & \text{при } \tau < 1, \qquad(17a)\\ 2a\tau\theta_c. & \text{при } \tau \geq 1. \qquad(17b) \end{cases}$$

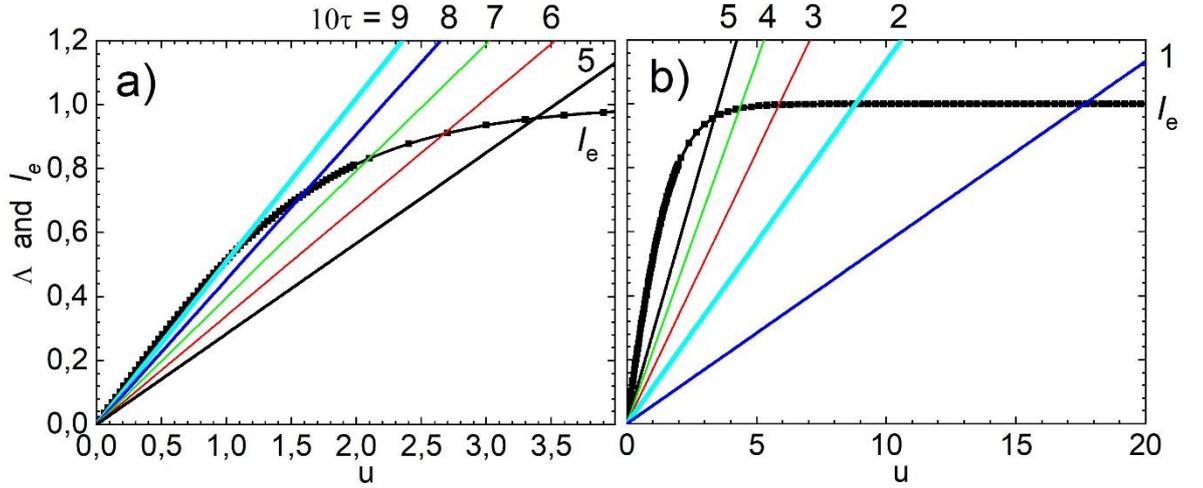

Fig. 5. Versus $u$ (see Eq. (16)) shown are right (exponential curve with saturation) and left parts of Eq. (14b). Each strait line corresponds to given value of $\tau$ which is indicated in the figure. The point of crossing provides a solution of Eq. (14b), i.e. the value of $\Lambda$. However, at small values of $\tau$ and near $\tau = 1$ finding these points numerically is significantly more difficult than calculating $\Lambda$ by means of the approximate Eqs. (14a) and (14c).

In Eqs. (17) $a$ denotes the single-particle level density parameter, $K_2(x)$ is the McDonald function (modified Bessel function) which we calculate as follows

$$K_m(x) = \int_0^{\infty} \text{ch}(mz)\exp(-x\,\text{ch}\,z)dz.\qquad(18)$$

The intrinsic energy $E$ is related to the entropy as:

$$E = \frac{S\theta}{2} + \frac{3a\Delta_0^2}{2\pi^2}(1 - \Lambda^2).\qquad(19)$$

In Fig. 6 we present the dependencies of the correlation function (see Eq. (14)), entropy (see Eq. (17)), and intrinsic energy (see Eq. (19)) upon the temperature obtained in our calculations. These results agree with the curves in Fig. 16 of Ref. [23] and with the Table of Ref. [28]. For our dynamical description of the capture process, we have to know the dependence of the temperature upon the intrinsic energy. We did not manage to find an analytical form for this dependence. That is why we use the tabulated values employed for creating the $E(\theta)$-dependence in Fig. 6. The obtained dependence $\theta(E)$ is shown in Fig. 7 and is used in further calculations. One sees in Fig. 7 that as a small value of $E$ appears, the temperature jumps up to approximately half of the critical value. After that the temperature rises smoothly.

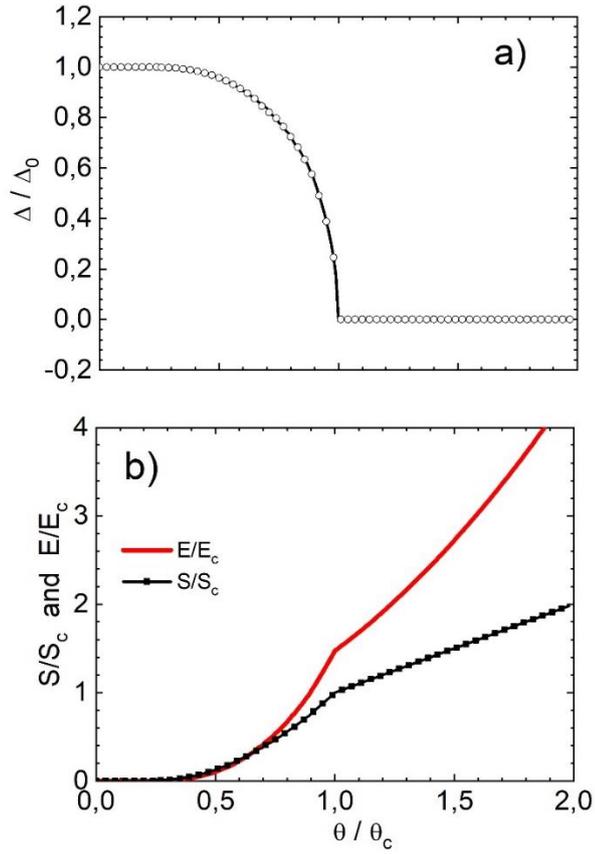

Fig. 6. The dependencies of the correlation function (a), as well as the entropy and intrinsic energy (b) upon the temperature obtained in our calculations.

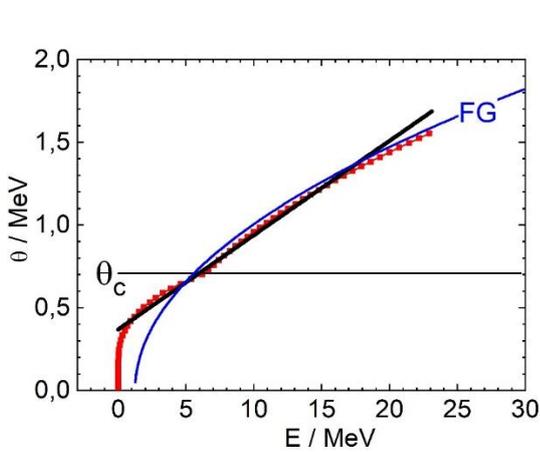

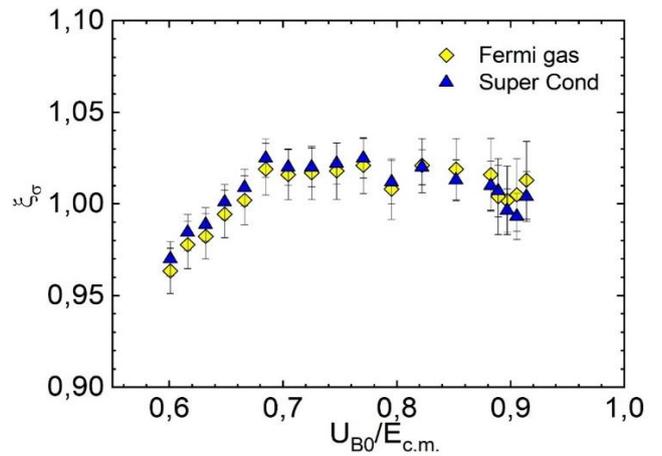

Fig. 7. Dependence of the temperature upon the intrinsic energy. Squares correspond to the calculations within the superfluid model; thick strait line represents the linear approximation of this calculation (see Eq. (20)); the calculations using the FGM are shown by the thin curve; horizontal strait line indicates the critical temperature. $a = 8.6\ MeV^{-1}$.

Fig. 8. The ratio of the calculated cross section $\sigma_{th}$ to the experimental one $\sigma_{exp}$ (see Eq. (7)) versus the ratio of the s-wave barrier energy to the collision energy. Diamonds stand for FGM calculations; triangles correspond to the calculations using the superfluid model. $K_R = 17\ zs/GeV$. The error bars correspond to the statistical errors of the calculations.

Dependence of the temperature upon the intrinsic energy during the dynamical modeling was approximated by the linear function:

$$\frac{\theta}{\theta_c} = 0.52 + 0.35 \frac{E}{a\theta_c^2}. \qquad (20)$$

Results of the dynamical modeling of the capture process are presented in Fig. 8 in the same way as in Fig. 4. Diamonds show results obtained within the FGM using Eq. (5) with $E_{DP} = 0$. Results of calculations within the superfluid model are presented by triangles. One sees that results of these two calculations are indistinguishable within 1% although from the theoretical point of view the superfluid model is preferable.

### 4. Conclusions

The dissipative character of the process of capture of two complex nuclei into orbital motion shows up as the conversion of the part of the collision kinetic energy into the intrinsic excitation energy of the colliding nuclei. This conversion is accounted for in different ways in some models. In the trajectory model with surface friction (TMSF) the dissipation strength and the intensity of fluctuations are related via the Einstein relation. The temperature of the system enters this relation, too. In all previous works (see, e.g., [4, 16, 19]) the temperature is related to the intrinsic excitation energy by means of the Fermi gas model (FGM).

Moreover, in that works the discrete structure of the intrinsic nuclear excitations was also ignored. This is justified for heavy or/and deformed nuclei but can be unrealistic for light or/and magic nuclei. Especially high is the energy of the first excited state of the oxygen-16 nucleus which is often used as the projectile nucleus in fusion experiments.

In the present work, we have studied the impact of the relation between the temperature and intrinsic excitation energy on the cross sections calculated within the framework of the TMSF. Reaction $^{16}$O+$^{92}$Zr has been considered as an example. We also have examined the suggestion that oxygen-16 nucleus probably stays in its ground state.

The results of the work can be formulated as follows:

i)Switching off the fluctuations completely, i.e., using zero temperature, results in 5% increase of the cross sections. Simultaneously the fusion excitation function becomes significantly oscillating.

ii)Using zero temperature for oxygen only results in 10% increase of the temperature for a given trajectory. However, when we calculate the fusion excitation function this increase does not show up and we unable to explain this.

iii)Using the superfluid model for the collision energies between 46 and 70 MeV results in the cross sections which are indistinguishable from the ones calculated by means of the FGM within the statistical errors of the modeling (typically about 2%).